\documentstyle[aps,eqsecnum,epsfig,rotating]{revtex}

\title{\hfill \begin{small} Preprint: WSU--NP--98--1 \end{small}\\
ANTI--LAMBDA/ANTI--PROTON RATIOS AT THE AGS\footnote{Submitted to {\em
    Proceedings of the 14th Winter Workshop on Nuclear Dynamics}, Snowbird,
  Utah, 31 January -- 6 February 1998,
  ``Advances in Nuclear Dynamics 4,'' W.\,Bauer and H.G.\,Ritter,
  eds. (Plenum Publishing, 1998).}}
\author{G.\,J.~Wang, R.~Bellwied, C.~Pruneau, and \underline{G.~Welke}}

\address{Department of Physics and Astronomy,\\ Wayne State University,
Detroit, MI 48202, U.S.A.\\
E--mail: welke@physics.wayne.edu}

\begin{document}
\maketitle

\begin{abstract}
  We attempt to explain the large ${\bar \Lambda}/\bar p$ ratios
  measured in heavy ion collisions at $12~{\rm A\cdot GeV}/c$ beam
  momentum within an hadronic framework.  The ratios are large
  compared to corresponding ratios in $pp$ collisions, and to thermal
  fits.  We show using a simple model and a detailed cascade
  calculation that different annihilation cross--sections of $\bar
  \Lambda$'s and $\bar p$'s, and the net conversion of $\bar p$'s to
  $\bar \Lambda$'s, do not account for the enhancement.  Uncertainties
  in elementary cross--sections and formation times are also
  considered.
\end{abstract}

\section{INTRODUCTION}

Anti--Lambda ($\bar \Lambda$) cross--sections in heavy ion collisions
are of interest because strangeness production is a potential signal
for QGP formation.\cite{shuryak,koch,rafelski}
 The ratio $\bar \Lambda/\bar p$ is of
special interest since it reflects the production of ${\bar s}$--quarks
relative to non--strange light anti--quarks, and should increase
substantially relative to the production expected from a superposition
of $NN$ collisions, if a QGP is formed.

Recent experiments have reported\cite{e859,e864b,na49,na35}
 measurements of $\bar p$
and $\bar \Lambda$ production in various heavy ion systems at the
Brookhaven AGS and at the CERN SPS. Experiment E859 reports the ratio
of $\bar \Lambda$ and $\bar p$ rapidity distributions to be $3 \pm 1
\pm 1$ in central Si+Pb collisions.\cite{e859} This ratio is corrected
for finite experiment acceptance, efficiencies, and for $\bar p$
creation from $\bar \Lambda$ decay (feed-down). It also takes into
account that neutral $\bar \Sigma$ particles cannot be distinguished
from the $\bar \Lambda$ sample.  Experiment E864 has
measured\cite{e864b} the $\bar p$ production cross--section in Au+Pb at
$11.6~{\rm A\cdot GeV}/c$.  They compare this measurement with a
similar one from the E878 collaboration that was obtained with a
focusing spectrometer, and interpret the difference between the two
measurements as an indicator of $\bar \Lambda$ production.  E864
estimates that $\bar \Lambda/\bar p > 2.3$ at the $98\%{\rm CI}$, with
a most probable value of 3.5. At the SPS, NA35 has
published\cite{na49,na35} $\bar \Lambda/\bar p$ ratios for $pp$, $pA$,
S+S, S+Ag, and S+Au collisions at $200~{\rm A\cdot GeV}/c$, and they
observe a significant rise from 0.25 for $pp$-collisions to 1.5 for
the heavy ion systems.

It is tempting to interpret the large reported ratios as evidence for
the formation of a QGP: Many authors\cite{letessier,let2,sollfrank}
 have argued
that the CERN multi-strange baryon ratios\cite{judd,dibari,kinson}
 can only be
described by a QGP scenario. This conclusion is, however, challenged
by E864 Au+Au RQMD simulations that find that the $\bar p$
production cross--section appears to be lower than expected from the
scaled $NN$ $\bar p$ cross--section.

Here, we shall try to address quantitatively all possible hadronic
contributions to the $\bar \Lambda/\bar p$ ratio.  We restrict
calculations to AGS energies and show that ``differential
annihilation'' of the two species and $\bar p$--to--$\bar \Lambda$
conversion processes can indeed enhance the $\bar \Lambda/\bar p$
ratio, but we conclude that the effect is not large enough -- thus
hinting at a production mechanism outside of the standard hadronic
interactions.  We present these arguments as follows: The observed
$\bar \Lambda/\bar p$ ratios cannot be explained by a thermal model,
unless severe inconsistencies with other measured data, such as the
charged pion and kaon cross--sections, are introduced. This has been
shown by many authors,\cite{pbm1,pbm2,heinz} and we give our own
calculation in the next Section. Thermal equilibration is unlikely at
present energies, so we consider transport simulations in the rest of
the paper, first discussing the relevant cross--sections, then the
results for a simple geometrical model, and finally the results of a
detailed cascade calculation.  Actual measurements of the $\bar
\Lambda$ and $\bar p$ cross--sections\cite{gjesdal,eisele} are used, not
event generator parametrizations.  We shall consider uncertainties in
these calculations, particularly those arising from the relatively
poorly known $\bar \Lambda$ annihilation, and from particle formation
times.  Speculative conclusions based on the quantitative discrepancy
between calculations and the actual measurements are presented in the
last Section.

\section{THERMAL MODEL}

In a thermal model, the final relative abundance of a particle is
determined by both the primary number of this species at freeze--out,
and by feed--down from heavier species after freeze--out.  We report
here results for central Au+Au collisions at $11.6~{\rm GeV}/c$, and
include all mesons with rest mass $\le 1~{\rm GeV}/c^2$, and all
baryons with rest mass $\le 1.7~{\rm GeV}/c^2$.

We assume thermal and chemical equilibration for freeze--out, so that
all relative abundances can be obtained from four parameters: the
freeze-out temperature $T_0$, the electric chemical potential $\mu_e$,
the baryonic chemical potential $\mu_b$, and the strange chemical
potential $\mu_s$.  Here,
\begin{equation}
 \mu_i \;=\; q_i\,\mu_e \:+\: b_i\,\mu_b \:+\: S_i \, \mu_s \nonumber 
\end{equation}
are particle chemical potentials, where $q_i$, $b_i$ and $S_i$ are the
charge, baryon number and strangeness of species $i$, respectively.
The thermal fit parameters are obtained by applying the conditions
$|Q/B-0.40| < 1\%$ and $|S|< 2\%$, together with the constraints
listed in Table~1.  We purposefully choose rather large ranges
in these latter constraints to show how difficult it would be to
obtain the experimental ${\bar \Lambda}/{\bar p}$ ratio.\footnote{Our
  number of $\bar \Lambda$'s throughout includes ${\bar \Sigma}^0$'s,
  as experiment does not distinguish between the two species.}

In Table~1, the ``errors'' on the best fit values for $T_0$ and the
chemical potentials are ranges that lead to results consistent with
the constraint intervals.\footnote{We note that these parameters are
  not inconsistent with overall energy conservation considerations.
  Recall also that $m_t$ spectra are blue shifted by the presence of
  strong flow.} We see that ${\bar \Lambda}/{\bar p}\, {\buildrel <
  \over \sim}\, 1.9$, at best.  It is instructive to consider the
results graphically.  Figure~1(a) shows ${\bar \Lambda}/{\bar p}$ as a
function of $K^{+}/K^{-}$ for various freeze--out temperatures $T_0$.
Large ${\bar \Lambda}/{\bar p}>2$ result only if the freeze--out
temperature and/or observed $K^+/K^-$ ratio are pushed unreasonably
high.  A similar conclusion follows from Figure~1(b), which shows
${\bar \Lambda}/{\bar p}$ as a function of $K^+/\pi^+$, with
$T_0=120\pm14~{\rm MeV}$.

We conclude that the experimental ratio ${\bar \Lambda}/{\bar p}$
ratio has, at the very least, a significant non--thermal component:
the lower experimental bound for the ratio reported by E864 is larger
than any reasonable thermal fit would allow.

\begin{small}
\begin{center}
\noindent TABLE 1 Thermal vs experimental particle ratios
for central Au+Au collisions at $11.6~{\rm GeV}/c$.
The parameter \\ranges are
$T_0= 120\pm14~{\rm MeV}$, $\mu_b=556\pm19~{\rm MeV}$,
$\mu_s= 111\pm14~{\rm MeV}$, and $\mu_e=-14\pm2~{\rm MeV}$.
\vskip 0.1in
\begin{tabular}{|c|c|c|ccc|}
\hline 
 &  & & \multicolumn{3}{c|}{Data} \\ \cline{4-6}
Ratio & Constraint  & Thermal Model & Ratio & Rapidity & Ref. \\ [0.5mm] \hline
$K^+$/$\pi^+$ & 0.16--0.28 &
$0.23\pm 0.03$ & $0.22\pm 0.01$ & 0.5--1.3 & \cite{Gonin} \\
$K^+/K^-$ & 4.0--6.0   & $4.73\pm 0.53$ & $5.0\pm 1.0$ &
0.5--1.3 & \cite{Gonin} \\
$K^-$/$\pi^-$& -- & $(3.50\pm 0.62)\times 10^{-2}$
        & 0.028 & 1.2--2.0 &\cite{Hiroyuki} \\ 
$\pi^+/p$ & 0.6--1.2   & $0.71\pm 0.09$& -- & -- & -- \\
$\pi^-/p$ & 0.8--1.4 & $1.00\pm 0.10$ & 1.00 & 1.2--2.0 &
\cite{Hiroyuki} \\
$\Lambda/p$ & -- & $0.16\pm 0.02$& -- & -- & -- \\
$\bar p/p$ & -- & $(3.48 \pm  3.44) \times 10^{-4}$ & -- & -- & -- \\
$\bar{\Lambda}/\bar{p}$ & -- & $1.58\pm 0.30$& -- & -- & -- \\ \hline
\end{tabular}
\end{center}
\end{small}

\section{CROSS--SECTIONS}

Generally, a thermal description is useful for a given particle
species if its mean free path is small compared to the system size.
This fact alone means that other approaches should also be
investigated. We do so in the next two sections, and discuss here the
most important physical input, {\em viz.}, the relevant
cross--sections.  The following are the relevant processes we shall
consider:  Firstly, the production of $\bar p$'s and $\bar \Lambda$'s:
\begin{eqnarray}
b_{1}+b_{2}&\rightarrow& N_{1}+N_{2}+B+\bar B~(+\pi)
\label{production_f}\\
M+b&\rightarrow& N+B+\bar B~(+\pi)
\label{production_2}\\
M+M&\rightarrow& B+\bar B~(+\pi)
\label{production_3}\\
S^{+}+S^{-}&\rightarrow& B+\bar B~(+\pi)
\label{production_4}\\
\end{eqnarray}

\begin{eqnarray}
b+S^{+}&\rightarrow& b+{\bar B}+N~(+{K^+}+\pi)
\label{production_5}\\
M+S^{+}&\rightarrow& N+{\bar B}~(+S^{+}+\pi) \label{production_l}
\end{eqnarray}
Here, $b$ represents a non-strange baryon, $N$ a nucleon, $B$ any
baryon, $M$ a light unflavored meson, and $S^{\pm}$ a $\pm 1$--strangeness
meson. Secondly, (same notation):
\begin{eqnarray}
\bar{p}+b &\rightarrow& X \label{annihilation_f}\\
b+\bar{\Lambda}&\rightarrow& S^{+}+\pi \label{annihilation_2}\\
S^{+}+\bar{b}&\rightarrow&{\bar \Lambda}+\pi\label{annihilation_3}\\
M+\bar{p}&\rightarrow&{\bar \Lambda}+\pi+S^{-} \label{annihilation_4}\\
M+\bar{\Lambda}&\rightarrow&{\bar b}+S^{+}(+\pi)\label{annihilation_l}
\end{eqnarray}
We refer to (\ref{annihilation_f}) and (\ref{annihilation_2}) as
annihilation processes, and (\ref{annihilation_3})--(\ref{annihilation_l}) 
as conversion processes.

The experimental $p\bar{p}$ annihilation cross--section
is well known, and can be para-\\ metrized as
\begin{eqnarray}
\sigma_{p\bar{p}}^{annih}(p_{lab})\;=\;67\:p_{lab}^{-0.7}~{\rm mb}~,
\label{PPbar}
\end{eqnarray}
where $p_{lab}$ is the momentum of the ``beam'' particle in ${\rm
  GeV}/c$ with the ``target'' at rest. The solid line in Figure~2
shows this parameterization of the data (diamonds).  
The $\bar{\Lambda}$ annihilation cross--section,
on the other hand, is relatively poorly
known, especially in the energy range we are interested in. We model
it by assuming that the elastic cross--sections for $p\bar{\Lambda}$
and $p\Lambda$ are equal, and then use the data of Ref.~\cite{eisele}
to obtain:
\begin{eqnarray}
\sigma_{p\bar{\Lambda}}^{annih}(p_{lab})\;=\;15\:
(p_{lab}/10)^{-\alpha}~{\rm mb}~, \label{PLbar_low}
\end{eqnarray}
where the same comments apply as for Eq.~(\ref{PPbar}).  While the
data is best fit by $\alpha=0.5$, the uncertainty is rather large (see
Figure~2; the stars are data from Ref.~\cite{eisele}).  In fact, one
might well argue that the data is consistent with

\begin{figure}
\begin{center}
\hspace{-1.5in}
\mbox{\epsfig{file=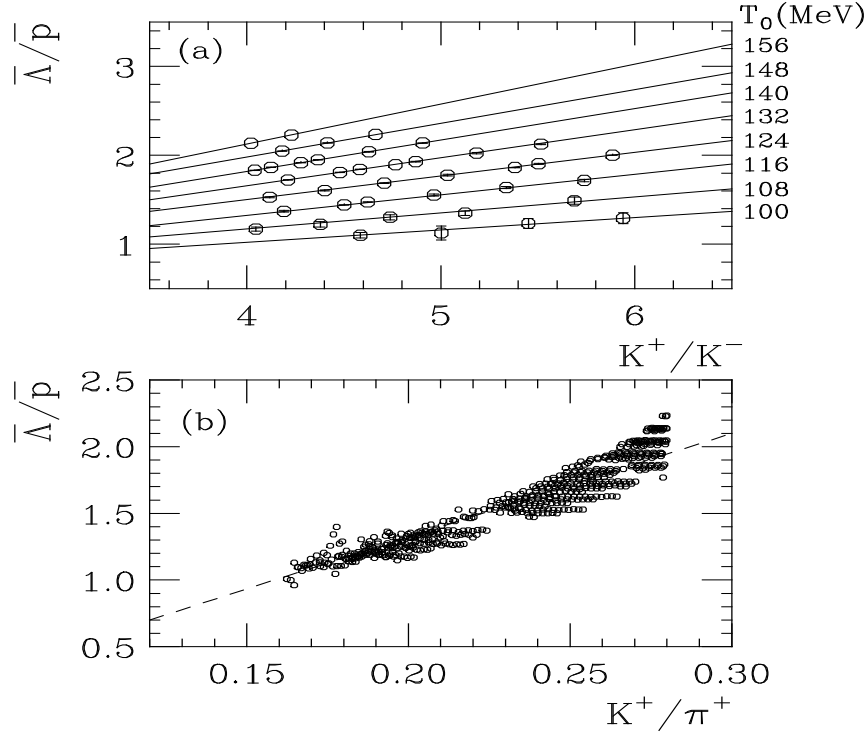,bbllx=221pt,bblly=358pt,bburx=540pt,
bbury=567pt,angle=90,width=5cm,height=6cm}}
\vspace{1.5in}
\caption{(a) Thermal ${\bar \Lambda}/{\bar p}$ ratios as a function
  of $K^{+}/K^{-}$ for various freeze--out temperatures $T_0$.  (b)
  Thermal ${\bar \Lambda}/{\bar p}$ ratios as a function of
  $K^+/\pi^+$, with $T_0=120\pm14~{\rm MeV}$. Error bars and scatter
points indicate values consistent with constraints not shown;
the lines are to guide the eye only.}
\end{center}
\end{figure}

\begin{figure}
\begin{center}
\hspace{-1.5in}
\mbox{\epsfig{file=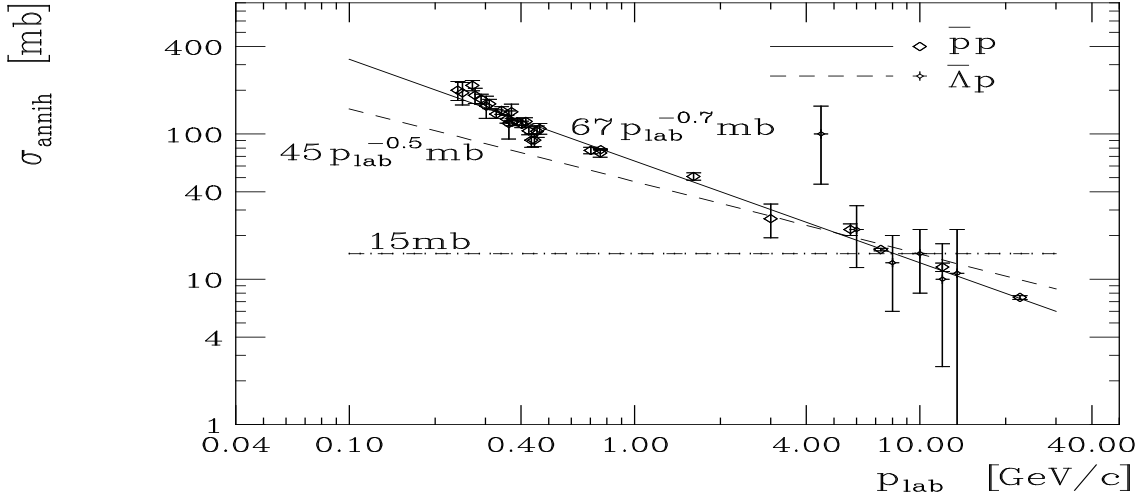,bbllx=221pt,bblly=358pt,bburx=540pt,
bbury=567pt,angle=90,width=5cm,height=5cm}}
\vspace{1.0in}
\end{center}
\caption{The annihilation cross--sections of
$\bar{p}$ ($\bar{\Lambda}$) with nucleons, as a function
of the incident momentum of the $\bar{p}$ ($\bar{\Lambda}$).}
\end{figure}

$\alpha=0$.\footnote{QCD sum rules imply a behavior consistent with
$\alpha \sim 0$.} For $\alpha=0.7$ the $\bar \Lambda$ data is
practically indistinguishable from the $\bar p$ data.  We shall
subsequently investigate the behavior of $\bar \Lambda/\bar p$ with
$\alpha$.

The $\bar{p}$ and $\bar{\Lambda}$ scattering with mesons is also an
important process we need to consider. Broadly speaking, we have three
types of collision: (1) thermalization of $\bar{p}$'s and
$\bar{\Lambda}$'s through elastic collisions; (2) production of
resonances that eventually decay back into $\bar{p}$'s or
$\bar{\Lambda}$'s; and (3), most importantly, net conversion of $\bar p$'s
to $\bar{\Lambda}$'s.  Chief amongst these is
\begin{equation}
\bar{p}+K^{+}\rightarrow \bar{\Lambda}~ ~ ~ ({\mbox {or resonances of }} 
\bar{\Lambda})\label{PKp}~ ~ ~,
\end{equation}
for which we know the charge conjugate reaction to have a sizeable
cross--section. The process (\ref{PKp}) thus contributes significantly
to reducing the $\bar{p}$ abundance while enhancing the
$\bar{\Lambda}$ abundance in the final state.  Given the pronounced
strangeness enhancement in large systems such as Au+Au, the process
(\ref{PKp}) should be a relatively important piece of the
$\bar{\Lambda}/\bar p$ ``puzzle.''

Finally, we shall need the ${\bar \Lambda}/{\bar p}$ ratio in $pp$
collisions. At $\sqrt{s}\sim 20~{\rm GeV}$, it has a
value\cite{blobel,antinuc,whit,GadRoh} 
of 0.25-0.30. At AGS energies ($\sqrt{s}\sim
5~{\rm GeV}$) its value is less established. Using
Refs.~\cite{blobel,antinuc,whit}, we infer a value of $\sim 0.2$, but
we shall use a value of 0.25 throughout this work. This implies that
the heavy ion ${\bar \Lambda}/{\bar p}$ ratios we obtain in subsequent
calculations will be upper bounds.

\section{GEOMETRIC MODEL CALCULATION}

It is often rather useful to have a simple understanding of cascade
output. We shall perform such a detailed cascade simulation in the
next Section; here we model $\bar p$ and $\bar \Lambda$ production,
annihilation and net conversion (${\bar p}$, ${\bar n}$ to ${\bar
  \Lambda}$, ${\bar \Sigma}$) in a simple geometric model. Our
starting point are the rate equations
\begin{eqnarray}
dN_{\bar{p}} &=& -(\frac{1}{\lambda_{a}}+\frac{1}{\lambda_{c}})\:
N_{\bar{p}}\:dz \nonumber\\
dN_{\bar{\Lambda}} &=& -\frac{1}{\Lambda_{a}}\:N_{\bar{\Lambda}}\:dz\:+\:
\frac{1}{\lambda_{c}}\:N_{\bar{p}}\:dz~,\label{rateeqn}
\end{eqnarray}
where $\lambda_a$ and $\lambda_c$ are the annihilation and net conversion
mean free paths for $\bar{p}$'s, respectively, and
$\Lambda_a$ is the annihilation mean free path for the
$\bar{\Lambda}$.  We take the initial ${\bar\Lambda}/\bar p$ ratio
from $pp$ collisions:
\begin{eqnarray*}
\bigg({\frac{N_{\bar \Lambda}}{N_{\bar p}}}\bigg)^{AA}{\bigg
  |}_{z=0}\;\equiv\;
\frac{N_{0}^{\bar p}}{N_{0}^{\bar \Lambda}}\;=\;
\bigg({\frac{N_{\bar \Lambda}}{N_{\bar p}}}\bigg)^{pp}\;
{\buildrel < \over \sim} \; 0.25~,
\end{eqnarray*}
as discussed in the previous section.  The final number of $\bar{p}$'s
or $\bar{\Lambda}$'s depends on the distribution of the length $z$ of
nuclear matter that the anti--particle passes through:
\begin{eqnarray*}
N\;=\;\langle N(z)\rangle~. \nonumber
\end{eqnarray*}
Integrating Eqs.~(\ref{rateeqn}) gives the number of $\bar{p}$ and
$\bar{\Lambda}$ after they have passed through a fixed length $z$ of
nuclear matter:
\begin{eqnarray}
N_{\bar{p}}(z) &=& N^{\bar p}_{0}\:{\rm e}^{-(1/\lambda_{a}+
1/\lambda_{c})z}
\label{Np_bar}\\
N_{\bar \Lambda}(z) &=& N^{\bar \Lambda}_{0}\:{\rm e}^{-z/\Lambda_{a}}
\:+\:2\,\frac{N^{\bar p}_{0}}{1+\frac{\lambda_{c}}{\lambda_{a}}-
\frac{\lambda_{c}}{\Lambda_{a}}}\:\bigg ( {\rm e}^{-z/\Lambda_{a}}-
{\rm e}^{-(1/\lambda_{a}+
1/\lambda_{c})z} \bigg )~.
\label{NL_bar}
\end{eqnarray}
The factor of two has been introduced to account for 
the net conversion of $\bar n$'s to ${\bar \Lambda}$'s.

Next, we need to model the geometry of the collision. We shall assume
that $z$ is related to the combined thickness of the beam and target
nuclei $t$ via
\begin{eqnarray*}
z\;=\;\beta\: t~ ~, \nonumber
\end{eqnarray*}
where $\beta<1$ is a constant. It measures the effect of matter
expansion, the local average momentum distribution, {\em etc}.
The distribution of $t$ in a central $AA$ collision is 
\begin{equation}
p(t)\;=\;\frac{dt^3}{(4R)^3}~ ~, \nonumber
\end{equation}
where $R$ is the nuclear radius. This equation allows us to compute
the survival probability $P_s=\langle {\rm e}^{-z/\lambda}\rangle$ of
a $\bar{p}$ or $\bar{\Lambda}$ as a function of $R$, $\lambda$ and
$\beta$. The analysis for $AB$ collisions is similar.

It remains to discuss the mean free paths. Generally,
\begin{equation}
\frac{1}{\lambda_j}\;=\;\sum_{i}\int\,dp_{lab}\:\rho_{i}(p_{lab})\:
\sigma_{ji}(p_{lab})\,\frac{dN_j}{dp_{lab}}~,
\end{equation}
where $p_{lab}$ is the momentum of the $\bar p$ or $\bar \Lambda$
measured in the rest frame of the target $i$ (nucleons, pions or
kaons). Also, $\rho_{i}$ is the density of target species $i$,
$\sigma_{ji}$ is the corresponding cross--section, and $dN_j/dp_{lab}$
is the distribution of $p_{lab}$.  Since the factor $\beta$ already
takes into account any changes in density, $\rho_i$ should be regarded
as the density for which no expansion occurs.

A rough approximation for $\lambda_{a}$ follows assuming that the
$\bar p$ and $\bar \Lambda$ are produced at rest in the $NN$ frame, and
are annihilated by the nucleons streaming by. In this case, $p_{lab}\sim
\sqrt{s/4-m_{p}^2} \sim 2.1~{\rm GeV}$, with $\rho\sim 0.16~{\rm
  fm^{-3}}$, and therefore $\lambda_{a} \sim 1.6~{\rm
  fm}$.\footnote{Note that $z/\lambda$ is Lorentz invariant; we
  perform calculations in the rest frame of the nucleus.} A more
detailed calculation that accounts for residual $\bar p$ motion in the
$NN$ frame gives $\lambda_{a}\sim 2.0~{\rm fm}$.  Similarly, using
appropriate cross--sections and thermal densities, we obtain
\begin{eqnarray*}
\Lambda_{a}&\approx& 4.2\: (3.76)^{-\alpha}~{\rm fm}\\
\lambda_{c}^{(1)}&\approx& 180~{\rm fm} \\
\lambda_{c}^{(2)}&\approx& 88~{\rm fm} \\
\lambda_{c}&\equiv& \frac{\lambda_{c}^{(1)}\lambda_{c}^{(2)}}
{\lambda_{c}^{(1)}+\lambda_{c}^{(2)}}\:\approx\: 59~{\rm fm}
\end{eqnarray*}
where $\lambda_{c}^{(1)}$ and $\lambda_{c}^{(2)}$ are the net
conversion mean free paths due to collisions with (N,$\pi$) and $K^+$,
respectively. 

The dependence of the two components of the ${\bar \Lambda}/
{\bar p}$ ratio ({\it viz.}, annihilation and conversion), 
are shown in Figure~3 for a central, symmetric ($AA$) collision,
as a function of the effective size of the system,
$R_{eff}\equiv \beta R$. The ratio 

\begin{figure}
\begin{center}
\hspace{-1.5in}
\mbox{\epsfig{file=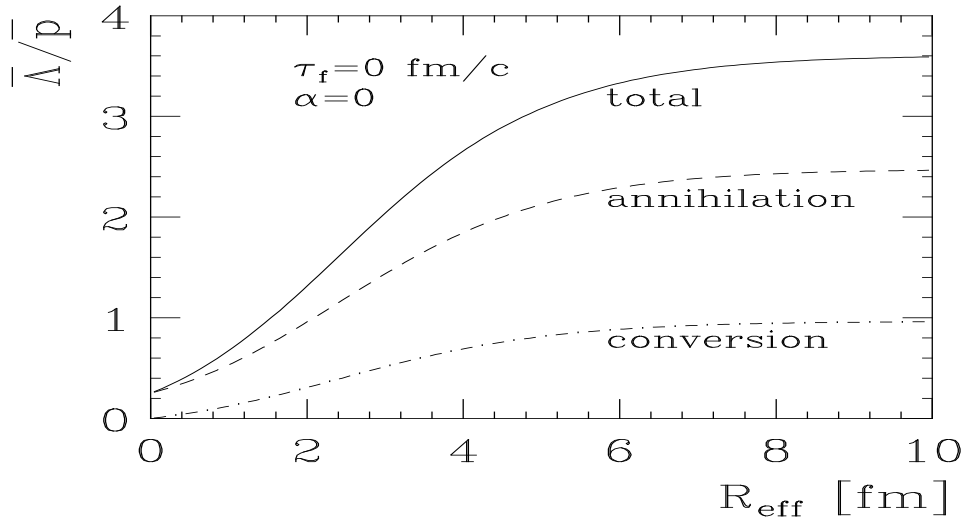,bbllx=221pt,bblly=358pt,bburx=540pt,
bbury=567pt,angle=90,width=5cm,height=5cm}}
\vspace{1.0in}
\end{center}
\caption{${\bar\Lambda}/\bar p$ as a function of $R_{eff}\equiv \beta
  R$ for a central $AA$ collision, with $\alpha=0$ and zero formation
  time (solid line). Individual contributions from annihilation and
  conversion are also shown.}
\end{figure}

saturates for large
$R_{eff}\,{\buildrel > \over \sim}\,6~{\rm fm}/c$  because
peripheral collisions play a increasingly important role; the ratio of
survival probabilities tends to the ratio of mean free paths cubed. 
On the other hand, for truly infinite matter all anti--particles pass
through a common large length, and ${\bar \Lambda}/{\bar p}$ is
exponentially increasing with $R_{eff}$.

The significant discrepancy between the data and our calculation (see
also next Section) may indicate that the ratio of core--produced $\bar
p$'s to peripherally produced ${\bar p}$'s is larger than predicted by
the simple binary hadron--hadron collision scenario.

The solid lines in Figure~4 show the final values of ${\bar
  \Lambda}/{\bar p}$ from this simple geometric model, if we use
$\beta=0.5$. The agreement with the cascade calculation (circles; see
next Section) is very good.

\section{CASCADE CALCULATION}

The production rate of $\bar p$'s or $\bar{\Lambda}$'s is only
$O(10^{-2})$ per event at the AGS, and cascade calculations are rather
CPU intensive. We shall therefore perform an effective calculation of
the survival probability by putting $\bar p$'s or $\bar{\Lambda}$'s in
by hand, wherever and whenever a collision occurs in which the energy
is sufficient for a $p\bar p$ or $\Lambda\bar{\Lambda}$ pair to be
produced. Thus we assume that the pair production does not depend on
energy, once above threshold.\footnote{Also, we assume that the
  production is not medium dependent.} The evolution of nucleons,
pions and kaons is not allowed to be influenced by the presence of
$\bar p$'s or $\bar{\Lambda}$'s -- we restore particles that
interacted with $\bar p$'s or $\bar{\Lambda}$'s to their
pre--collision kinematics.

We may then calculate the ${\bar \Lambda}/{\bar
  p}$ ratio via the survival probabilities $P_s$
\begin{equation} 
\frac{\bar \Lambda}{\bar p}\;=\;r_p\,\frac{P_s({\bar \Lambda})}
{P_s({\bar p})}\:+\:2 r_c\,\frac{{\bar \Lambda}^c}{{\bar p}^s}~ ~,
\label{final_ratio}
\end{equation} 
where ${\bar p}^s$ is the number of surviving ${\bar p}$'s and ${\bar
  \Lambda}^c$ the number of converted ${\bar \Lambda}$'s. Also,
$r_p\sim 0.25\pm 0.1$ is the ${\bar \Lambda}/{\bar p}$ ratio in $pp$
collisions, while $r_c\approx 0.25$ is a correction factor.\footnote{We
overestimate $\bar \Lambda$--production because we assume that all
collisions of non-strange anti--baryons with positive strange mesons
result in a ${\bar \Lambda}$.} The factor two accounts for the
conversion of $\bar n$'s to ${\bar \Lambda}$'s.

To illustrate the effect of differential annihilation versus net
conversion, we show in Table~2 the different components of
Eq.~(\ref{final_ratio}) for different $\alpha$ and $\tau=0$ in central
Au+Au collisions at AGS energies.  For small $\alpha$ (small ${\bar
  \Lambda}$ annihilation cross--section), ${\bar \Lambda}/{\bar p}$
enhancement results almost entirely from differential annihilation.
As $\alpha$ increases, the survival probability of a ${\bar \Lambda}$
and a ${\bar p}$ become equal, and the sole enhancement in the final
ratio results from conversion.

\begin{small}
\begin{center}
\noindent TABLE 2 Survival probabilities, the conversion rate, and final
  ${\bar\Lambda}/{\bar p}$\\ ratio for various $\alpha$ and $\tau=0~{\rm
    fm}/c$, in a central Au+Au collision.
\vskip 0.1in
\begin{tabular}{|c|ccc|c|} \hline
   \raisebox{-0.1cm}{$\alpha$} & \raisebox{-0.1cm}{$P_{s}({\bar\Lambda})$
    (\%)} & \raisebox{-0.1cm}{$P_{s}({\bar p})$ (\%)} &
    \raisebox{-0.1cm}{${\bar\Lambda}^{c}/{\bar p}^s$}&
    \raisebox{-0.1cm}{${\bar\Lambda}/{\bar p}$} \\
    \hline 0.0 & $12.1\pm 0.3$ & $1.5\pm
    0.1$ & $0.75\pm0.07$ & $2.4\pm 1.0$ \\ 0.2 & $6.0\pm 0.2$ &
    $1.5\pm0.1$ & $0.45\pm0.06$ & $1.2\pm0.5$ \\ 0.4 & $3.3\pm0.2$ &
    $1.5\pm 0.1$ & $0.38\pm0.06$ & $0.7\pm0.3$ \\ 0.7 & $1.5\pm0.05$ &
    $1.5\pm0.1$ & $0.16\pm0.02$ & $0.3\pm0.1$ \\ \hline
\end{tabular}
\end{center}
\end{small}

\begin{figure}
\begin{center}
\hspace{-1.5in}
\mbox{\epsfig{file=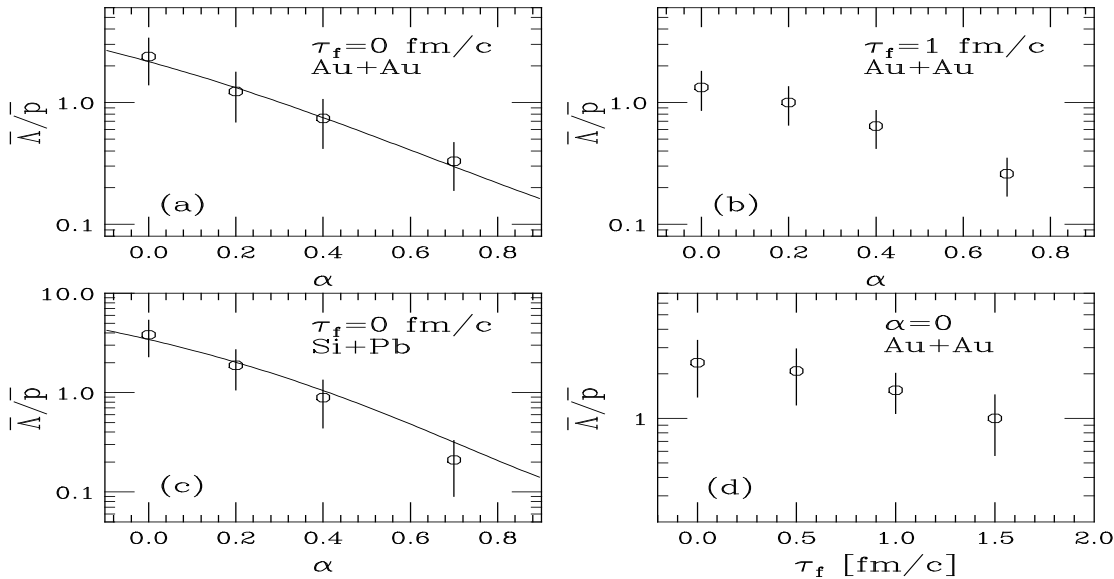,bbllx=221pt,bblly=358pt,bburx=540pt,
bbury=567pt,angle=90,width=5cm,height=5cm}}
\vspace{1.2in}
\end{center}
\caption{Cascade ${\bar \Lambda}/{\bar p}$ ratios for various
  $\alpha$, $\tau$ and systems. The solid lines are for the 
  geometric model with $\beta=0.5$.}
\end{figure}

We show in Figure~4 the final ${\bar\Lambda}/{\bar p}$ ratios for
several values of $\alpha$ and $\tau$ for various systems.  The solid
lines are results from our simple geometric model using $\beta=0.5$,
and show that the simple geometrical model predicts the
ratios rather well. We note that the ratio is larger in Si+Pb
collisions than in Au+Au collisions, both in our calculation and in
experiment. Also, note that (1) realistic values for $\alpha$ are probably
closer to 0.5 than 0.0; (2) formation times are 
probably not zero; and (3)
our initial ({\it i.e.}, $pp$) ${\bar \Lambda}/{\bar p}$
ratio is probably an over--estimate. We thus
conclude that in our present calculation ${\bar\Lambda}/{\bar p}\sim
1$ and most definitely ${\buildrel < \over \sim}\, 2$ for the Au+Au
system (${\buildrel < \over \sim}\, 3$ for Si+Pb).

\section{CONCLUSIONS}

The ${\bar \Lambda}/{\bar p}$ ratio in $NN$ collisions at AGS energies
is ${\buildrel < \over \sim}\, 0.25$, heavy ion experiments measure
much larger values.  The ratio can only be pushed to values above 2
with much difficulty in a hadronic thermal model with chemical
equilibration. For Au+Au, using upper limits of $K^{+}/\pi^{+}=0.30$,
$K^{+}/K^{-}=6.0$, and $T_0=140~{\rm MeV}$, we obtain an upper limit
of ${\bar \Lambda}/{\bar p} \sim 2$, below the E864 98\%--lower
confidence limit of 2.3.

We have considered a non--equilibrium description, and presented
results from a simple geometric model and a detailed cascade
calculation. The geometric model does rather well in reproducing
trends in the data.  Chief inputs to our calculations are the ${\bar
  \Lambda}$ and ${\bar p}$ annihilation cross--sections with nucleons,
as well as processes that convert ${\bar p}$'s into ${\bar
  \Lambda}$'s.  The ${\bar \Lambda}$ annihilation cross--section is
relatively poorly known, but it may be argued that it is somewhat less
than ${\bar p}$ annihilation at AGS energies. Our parameter $\alpha$
controls this difference, with $\alpha=0$ corresponding to the
greatest difference, and $\alpha=0.7$ corresponding to a
cross--section that is practically indistinguishable from ${\bar p}$
annihilation. Currently, the cross--section data is best fit by
$\alpha=0.5$.

In Au+Au cascade simulations at AGS energies, the largest ${\bar
  \Lambda}/{\bar p}$ ratio obtainable is $\sim 3.5$, for $\alpha=0$
and $\tau=0~{\rm fm}/c$. Both these input values are extreme. What we
currently believe to be more reasonable inputs lead to a value of
${\bar \Lambda}/{\bar p} \sim 1$, far from the lower bound of the E864
experiment.

Interestingly, higher ${\bar \Lambda}/\bar p$ ratios result in central
Si+Pb collisions, both in our calculations and in experiment.  Within
our model this has a simple geometric explanation: For Si+Pb, the
$\bar p$'s and $\bar \Lambda$'s are produced close to the beam axis,
leading to a larger average nuclear thickness that the anti--particle
must traverse.  We also note that at higher energies the
difference between the ${\bar \Lambda}$ and ${\bar p}$ annihilation
cross--section practically vanishes. One might therefore predict (in
hindsight) that ${\bar \Lambda}/{\bar p}$ at the SPS should be lower
than at the AGS, as indeed observed.

We conclude that the large ${\bar \Lambda}/\bar p$ ratios in AGS heavy
ion collisions are not easily explained by hadronic mechanisms.  QGP
formation might be a possible solution, but a more accurate
measurement of the $\bar \Lambda$ annihilation cross-section in the
relevant energy range is sorely needed before any more definite
conclusion can be reached.

This work was supported in part by the U.S. Department of Energy
under Grant No. DE-FG02-93ER40713.

\end{document}